\documentstyle{article}
\title{{\bf A local realist theory of parametric down
conversion}}
\author{Trevor~W.~Marshall\\
Dept. of Mathematics, University of Manchester,\\
Manchester M13 9PL, U.K.}
\date{\today}
\begin{document}
\maketitle
\begin{abstract}
In a series of articles we have
shown that all parametric-down-conversion processes, both of type-I and
type-II, may be described by a positive Wigner density. These results,
together with our description of how light detectors subtract the
zeropoint radiation, indicated the possibility of a completely local realist
description of all these processes. In the present article we show how
the down-converted fields may be described as retarded
fields generated by currents inside the nonlinear crystal,
thereby achieving such a theory.
Most of its predictions coincide with the standard
nonlocal theory. However, the intensities of the
down converted signals do not correspond exactly with
the photon pairs of the nonlocal theory.
For example, in a blue-red down conversion we would
find about 1.03 red "photons" for every blue one.
The theory also predicts a new
phenomenon, namely parametric up conversion
from the vacuum.
\end{abstract}
\section{Introduction}
We have treated type-I parametric down conversion (PDC) processes,
in the Wigner formalism, in a series of articles\cite{pdc1,pdc2,pdc3},
and we extended this treatment to type-II PDC in \cite{pdc4}. In all of
these processes, the resulting Wigner density is positive, as is, rather
trivially, that of the vacuum. We have also proposed, in these articles,
a theory of detection which is formally almost identical with the
standard normal-ordering prescription of quantum optics. However, our
description of the detection process recognizes that the vacuum
fluctuations are real, so an important element of the theory is the
manner in which detectors are able to extract signals from the rather
large zeropoint noise background. This problem was discussed in
\cite{pdc4}, and we indicated the way towards its solution.

The approach of the above series of articles
was a kind of compromise between the standard,
nonlocal theory of Quantum Optics, where the
interaction of the various field modes is
represented by a hamiltonian, and a fully
maxwellian theory, which would be both
local and causal. In this latter case, the
nonlinear crystal would be represented as
a spatially localized current distribution,
modified of course by the incoming
electromagnetic field; the outgoing field
would then be expressed as the retarded
field radiated by this distribution.
A preliminary attempt at such a theory was made\cite{magic},
using first-order perturbation theory.
However, we showed, in the above series of articles, that a calculation
of the relevant counting rates, to lowest order, requires us to find the
{\it second}-order perturbation corrections to the Wigner density,
and the close formal parallel between these two theories
means that the same considerations will apply to
the maxwellian theory.

If we
were to take account of the tensor
character of the polarizabilities, this would represent a rather
formidable task. In this article we study a simplified model of the
crystal, in which the electric field,
and hence also the linear and nonlinear polarizabilities, are
considered to be scalars.   
In such a model it is not possible to
discuss the polarization correlation of the
signal with the idler, so we are reverting to
type-I PDC, which means we confine attention to
the frequency and angular
correlations between these two beams.

We shall also make the simplifying assumption
that the crystal is infinitely large in the directions perpendicular to
the pumping beam. This reduces the problem to a single spatial
dimension, and, after making a certain
linearization approximation,
allows us to pass to a nonperturbative treatment of the
process. We shall simplify the algebra by assuming a constant value
for the nonlinear polarizability, but it will nevertheless be essential
to retain an explicit frequency dependence for the
linear part of the polarizability.
\section{The linearization procedure}
Provided the pumping laser is
sufficiently intense and coherent, it is possible
to neglect the depletion in its intensity which
occurs when it interacts with other modes
of the light field. This leads to a
linearization of the field equation inside
the nonlinear crystal. We remark that this
procedure is essentially the same as is used
in the standard photon analysis, where, by
treating the laser amplitude as a $c$-number,
a cubic interaction term in the hamiltonian
is reduced to a quadratic.

The scalar electric field $E(x,y,z,t)$ satisfies the
wave equation
\begin{equation}
\Delta E-\ddot E=0
\end{equation}
outside the crystal, and
\begin{equation}
\Delta E-\ddot E=-4\pi\dot J\;
\end{equation}
inside the crystal, where $J(x,y,z,t)$ is the current.
The relation connecting $J$ with $E$ is
\begin{equation}
J({\bf r},t)= \frac{1}{2\pi}\int_{-\infty}^\infty
d\omega\int_{-\infty}^\infty dt'
i\omega f(\omega)e^{i\omega t-i\omega t'}E({\bf r},t')+
g'[E({\bf r},t)]^2\;,
\end{equation}
where  $f(\omega)$ is analytic in the lower half plane,
so that the integration on $t'$ may be taken from
minus infinity to $t$ only. The refractive index
is then  given by
\begin{equation}
\mu^2(\omega)=1+4\pi f(\omega)\;.
\end{equation}
The Fourier transformed
wave equation, inside the crystal, is
\begin{equation}
\Delta\tilde E(\omega)-\omega^2\mu^2(\omega)\tilde E(\omega)=
-4\pi i\omega g'\int E^2(t)e^{i\omega t}dt\;.
\end{equation}
We shall suppose that the laser field inside
the crystal is
\begin{equation}
E_L(x,y,z,t)=V\cos[\omega_0\mu(\omega_0)z-\omega_0t]\,,
\end{equation}
which represents a plane wave travelling from
left to right. We should include a right-to-left
wave resulting from internal reflection of the
laser, but, in the linearization approximation
we are about to describe, such a wave simply
produces additional but independent pairs of down-converted
signals. These are easily treated by our
method, but for simplicity we shall not
include them.
Our linearization now consists of putting $E=E_L+E'$, and
discarding terms in $E'^2$. Provided we do not have
$\omega$ close to a multiple of $\omega_0$, we may also
discard the terms in $E_L^2$, so our linearized wave equation,
putting $g=4\pi g'V/\omega_0$, is
\begin{eqnarray}
\Delta\tilde E'(\omega)-\omega^2\mu^2(\omega)\tilde E'(\omega)=&
-ig\omega\omega_0[
\tilde E'(\omega-\omega_0)e^{i\omega_0\mu(\omega_0)z}
\nonumber\\
&+\tilde E'(\omega+\omega_0)e^{-i\omega_0\mu(\omega_0)z}]\;.
\label{scalmax}
\end{eqnarray}
\section{What is PDC?}
It is necessary to pose this question, because,
depending on the answer given, PDC may be
described as either a local or a nonlocal
phenomenon.
This is because the view we are advocating
requires us to recognize the reality of the
zeropoint electromagnetic field (ZPF). The
connection between a real ZPF and locality,
in the context of PDC, was discussed in
the last article of the Wigner series\cite{pdc4}.

An example of the modern, nonlocal description
is provided by Greenberger, Horne and Zeilinger\cite{ghz}.
A nonlinear crystal (NLC), pumped by a laser at frequency
$\omega_0$, produces conjugate pairs of signals,
of frequency $\omega$ and $\omega_0-\omega$ (see Fig.1).
Since the modern wisdom is that light consists
of photons, this means that an incoming laser
photon ``down converts" into a pair of lower-energy
photons. Naturally, since we know that $E=\hbar\omega$,
that means energy is conserved in the PDC process,
which must be very comforting.

There is an older description, which I suggest is
more correct than the modern one. It had only a short life.
Nonlinear optics was born in the late 1950s, with the
invention of the laser, and, up to about 1965, when
Quantum Optics was born, the PDC process would
have been depicted\cite{saleh} by Fig.2; an incoming wave
of frequency $\omega$ is down converted, by
the pumped crystal, into an outgoing signal
of frequency $\omega_0-\omega$. The explanation
of the frequency relationships lies in the
multiplication, by the nonlinear crystal,
of the two input amplitudes; we have no need of $\hbar$!

This process persists when the intensity
of the input is reduced to zero. This is because
all modes of the light field are still
present in the vacuum, and the nonlinear crystal
modifies vacuum modes in exactly the same way
as it modifies input modes supplied by an
experimenter.
What we see  emerging from the crystal is
the familiar PDC rainbow.
This is because the angle of incidence
$\theta$, at which PDC
occurs, is different for different frequencies,
on account of the variation of refractive
index with frequency. We depict the
process of PDC from the vacuum in Fig.3,
but note that this figure shows only two
conjugate modes of the light field; a complete
picture would show all frequencies participating
in conjugate pairs, with varying angles
of incidence. In contrast with Fig.2, where we
showed only the one relevant input, we must now
take account also of the conjugate input mode
of the zeropoint, since the first mode itself has
only the zeropoint amplitude.
The zeropoint inputs, denoted by
interrupted lines in
Fig.3, do not activate photodetectors, because the
threshold of these devices is set precisely at the
level of the zeropoint intensity, as discussed in ref\cite{pdc4}.
However, as we shall see in the next Section,
the two idlers have intensities above
that of their corresponding inputs.
Also there is no
coherence between a signal and an idler of the same
frequency, so  their intensities
are additive in both channels. Hence there are photoelectron
counts in both of the outgoing channels of Fig.3.

The question we have posed in this section could
be rephrased as ``What is it that is down converted?".
According to the thinking behind Fig.1, the laser
photons are down converted, whereas according to
Fig.3 it is the zeropoint modes; they undergo both
down conversion, to give signals, and amplification,
to give idlers.

\section{The down-conversion process}\label{pdcsect}
We begin by reviewing the standard
treatment of linear dielectric laminas, in
the approximate form described in Freedman's
thesis\cite{freedman} and used by many
since, including Aspect\cite{aspect} and ourselves\cite{found}.
The approximation holds if the thickness $l$ of the
lamina is large compared with the wavelength,
and ignores interference effects between successive
internal reflections. The lamina would need to be
extremely accurately cut, in any case, for such
effects to be observed.
Let the lamina occupy the region $0<z<l$.
A plane scalar wave is represented by
\begin{equation}
E(x,y,z,t)=e^{ip_x x+ip_y y+i\Omega_0(p)z-i\omega t}\;,
\end{equation}
where
\begin{equation}
\Omega_0^2(p)=\omega^2-p^2\quad{\rm and}\quad p^2=p_x^2+p_y^2.
\end{equation}
We shall put
\begin{equation}
E(x,y,z,t)=F(z,t)e^{ip_x x+ip_y y}.
\end{equation}
Then, for $g=0$, a solution of eq.(\ref{scalmax}) may be
found in the form
\begin{eqnarray}
F(z,t)&=&e^{-i\omega t}(e^{i\Omega_0 z}+Re^{-i\Omega_0 z})(z<0)\;,
\nonumber\\
F(z,t)&=&e^{-i\omega t}(Ae^{i\Omega z}+Be^{-i\Omega z})(0<z<l)\;,
\nonumber\\
F(z,t)&=&e^{-i\omega t}Te^{i\Omega_0 z}(z>l)\;,
\end{eqnarray}
where
\begin{equation}
\Omega^2(p)=\omega^2\mu^2(\omega)-p^2\;.
\end{equation}
The four constants $(R,A,B,T)$ may be determined
by imposing the conditions that $F$ and
$\partial F/\partial z$ be continuous at
$z=0$ and $z=l$. A solution procedure
consists of first putting $B=0$ and
imposing the boundary conditions at $z=0$ only,
thereby obtaining values for $R$ and $A$;
then we impose the boundary conditions at
$z=l$, using the value previously obtained for $A$,
to obtain $B$ and $T$; then we go back to $z=0$
with the new value of $B$ and calculate corrected
values for $R$ and $A$; and so on. If we were
to sum the infinite series, this would give us
an exact solution.
The first step gives
\begin{eqnarray}
R_0 &=&\frac{\Omega_0-\Omega}{\Omega_0+\Omega}\;,
\nonumber\\
A_0 &=&\frac{2\Omega_0}{\Omega_0+\Omega}\;.
\end{eqnarray}
The corresponding intensity coefficients are
obtained by considering the ratios of the
Poynting vector's $z$-component, that is
$E\partial E/\partial z$. They are
\begin{equation}
r_0=R_0^2\qquad{\rm and}\qquad t_0=A_0^2\Omega/\Omega_0=1-r_0\;.
\end{equation}
Note that we are here neglecting the imaginary
part of $\mu$, which gives rise to a small
absorption rate.
The second step gives, apart from a phase factor which
we do not need,
\begin{equation}
B_0=A_0R_0\qquad{\rm and}\qquad T_0=A_0^2\Omega/\Omega_0\;,
\end{equation}
which again gives the intensity coefficients $r_0$ and $t_0$.
Our approximation consists of simply multiplying
by the appropriate intensity factors for all
internal reflections and transmissions, thereby
arriving at the overall coefficients
\begin{eqnarray}
r&=&r_0+r_0t_0^2+r_0^3t_0^2+\ldots=\frac{2r_0}{1+r_0}\;,
\nonumber\\
t&=&t_0^2+t_0^2r_0^2+t_0^2r_0^4+\ldots
=\frac{1-r_0}{1+r_0}\;.
\end{eqnarray}
These, naturally, are independendent of the lamina
thickness, and satisfy $r+t=1$. Of course, in
this case of $g=0$, the above result is completely
unaffected by the presence of a pumping laser.

Now we shall extend this theory to the nonlinear
dielectric, described in the previous section.
We put
\begin{equation}
F(z,t)=\tilde F(z,\omega)\,e^{-i\omega t}
+\tilde F(z,\omega-\omega_0)\,e^{i\omega_0 t-i\omega t}\;.
\end{equation}
Then a solution of eq.(\ref{scalmax}) is
\footnote{Here we make the approximation of
discarding the coupling with waves of frequency
$\omega+\omega_0$. This is justified because
we are assuming that $\omega$ and $p$ are close
to the PDC resonance condition given by eq.(\ref{rainbowd}),
and are far from the corresponding PUC resonance
condition. In section \ref{puc} we shall be assuming
that the reverse is the case.
}
\begin{eqnarray}
\tilde F(z,\omega)&=&\sum_{r=1}^4\alpha_r\,e^{ik_r z}\,,
\nonumber\\
\tilde F(z,\omega-\omega_0)&=&i\sum_{r=1}^4
\beta_r\,e^{i[k_r-\omega_0 \mu(\omega_0)]z}\,,
\label{intfield}
\end{eqnarray}
where
\begin{eqnarray}
\quad[k_r^2-\Omega_1^2(p)]\alpha_r&=&g\omega_0\omega
\beta_r\,,
\nonumber\\
\quad[\{k_r-\omega_0 \mu(\omega_0)\}^2 -\Omega_2^2(p)] \beta_r&=&
g\omega_0(\omega_0-\omega)
\alpha_r\,,\label{modecoup}
\end{eqnarray}
and
\begin{eqnarray}
\Omega_1(p)&=&+\sqrt{\omega^2\mu^2(\omega)-p^2}\;,\nonumber
\nonumber\\
\Omega_2(p)&=&
+\sqrt{(\omega_0-\omega)^2\mu^2(\omega_0-\omega)-p^2}\;.
\end{eqnarray}
We define also the corresponding free-space quantities
\begin{eqnarray}
\Omega_{10}(p)&=&+\sqrt{\omega^2-p^2}\;,\nonumber\\
\Omega_{20}(p)&=&
+\sqrt{(\omega_0-\omega)^2-p^2}\;.
\end{eqnarray}

We now put
\begin{eqnarray}
k_1&=&\Omega_1+\epsilon_1\,,\nonumber\\
k_2&=&\Omega_1+\epsilon_2\,,\nonumber\\
k_3&=&-\Omega_1+\epsilon_3 \,,\nonumber\\
k_4&=&\Omega_2+\omega_0 \mu(\omega_0) +\epsilon_4\,. \label{kval}
\end{eqnarray}
There is a particular value of $p$, which we shall
designate $p_0$, satisfying the condition
\begin{equation}
\Omega_1(p_0)+\Omega_2(p_0)=\omega_0\mu(\omega_0)\;.\label{rainbowd}
\end{equation}
This defines the direction $\theta(\omega)$
(see Fig.2) of the zeropoint field
component, of frequency $\omega$, for which the
PDC ``resonance'' has its maximum. We denote
\begin{equation}
\omega_1=\Omega_1(p_0)\quad,\quad\omega_2=\Omega_2(p_0)\;,
\end{equation}
and define $\omega_{10}$ and $\omega_{20}$ similarly.
Then, for $g<<1$ and $\mid p-p_0\mid<<\omega$,
$\epsilon_r$ are given by
\begin{eqnarray}
\epsilon_1+\epsilon_2&=&(p-p_0)\frac{p_0(\omega_1+\omega_2)}
{\omega_1\omega_2}\;,\label{epssum}\\
\epsilon_1\epsilon_2&=&\frac{g^2\omega(\omega_0-\omega)\omega_0^2}
{4\omega_1\omega_2}
\;,\label{epsprod}\\
\epsilon_3&=&-\frac{{g^2\omega(\omega_0-\omega)\omega_0^2}}
{8(\omega_1+\omega_2)\omega_1^2}
\;,\\
\epsilon_4&=&\frac{{g^2\omega(\omega_0-\omega)\omega_0^2}}
{8(\omega_1+\omega_2)\omega_2^2}
\;.
\end{eqnarray}

Because of the smallness of
$\epsilon_2-\epsilon_1$, the modes $k_1$ and $k_2$
are strongly coupled, and the phase relations
between them, at both the frequencies $\omega$
and $\omega_0-\omega$, are carried from one side of the crystal to the
other.
These modes all represent left-to-right waves.
On the other hand, the modes associated with $k_3$
and $k_4$ represent right-to-left waves. Because of the
smallness of $\epsilon_3$ and $\epsilon_4$,
$k_3$ and $k_4$ are well separated, and
these two waves are effectively
unaltered by the interaction with the laser, which
is explained by the fact that the laser wave
travels in the opposite direction to them. We therefore put
\begin{equation}
\alpha_1=A_1\;,\;\alpha_2=A_2\;,\;\alpha_3=A_3\;,\;\beta_4=A_4\,,
\end{equation}
and then, neglecting $\epsilon_3$ and $\epsilon_4$, the
internal field (\ref{intfield}) becomes
\begin{eqnarray}
F(z,t)=A_1\left[e^{i(\Omega_1+\epsilon_1)z-i\omega t}
+\frac{2\omega_1\epsilon_1}{g\omega\omega_0}e^
{-i(\omega_0\mu(\omega_0)-\Omega_1
-\epsilon_1)z+i(\omega_0-\omega)t}\right]\nonumber\\
+A_2\left[e^{i(\Omega_1+\epsilon_2)z-i\omega t}
+\frac{2\omega_1\epsilon_2}{g\omega\omega_0}e^
{-i(\omega_0\mu(\omega_0)-\Omega_1
-\epsilon_2)z+i(\omega_0-\omega)t}\right]\nonumber\\
+A_3e^{-i\Omega_1 z-i\omega t}
+A_4e^{i\Omega_2 z+i(\omega_0-\omega)t}\;.
\end{eqnarray}
This must be matched with
\begin{equation}
F(z,t)=e^{i\Omega_{10} z-i\omega t}+R_1 e^{-i\Omega_{10} z-i\omega t}
+R_2 e^{i\Omega_{20} z + i(\omega_0-\omega)t}
\end{equation}
for $z<0$, and
\begin{equation}
F(z,t)=T_1 e^{i\Omega_{10} z-i\omega t}+
T_2 e^{-i\Omega_{20} z + i(\omega_0-\omega)t}
\end{equation}
for $z>l$.
We must impose four boundary conditions (continuity
of $F$ and of $\partial F/\partial z$ for both frequencies) at
$z=0$, and a similar four at $z=l$. This will
determine the eight constants $(R_1,R_2,T_1,T_2,A_1,
A_2,A_3,A_4)$. Our iteration procedure consists
in putting $A_3=A_4=0$, and then solving for $(R_1,R_2,
A_1,A_2)$ using only the boundary conditions at $z=0$;
then, using these values for $A_1$ and $A_2$, we
solve for $(A_3,A_4,T_1,T_2)$ using the boundary
conditions at $z=l$; then, using these values of
$A_3$ and $A_4$, we calculate the
corrections to $(R_1,R_2,A_1,A_2)$ using the
boundary conditions at $z=0$; and so on.
This series of iterations is greatly simplified
by making the same approximation as we made
above in the linear case, that is to say by
neglecting interference effects between
successive reflections inside the crystal,
with the exception that in this case, for the
reasons given above, we cannot neglect
interference between $A_1$ and $A_2$.

The first step of the above procedure leads to the
four equations
\begin{eqnarray}
1+R_1&=&A_1+A_2\,,
\nonumber\\
\omega_{10}(1-R_1)&=&(\omega_1+\epsilon_1)A_1+(\omega_1+\epsilon_2)A_2\,,
\nonumber\\
R_2&=&\frac{2\omega_1}{g\omega\omega_0}(\epsilon_1 A_1+
\epsilon_2A_2)
\,,\nonumber\\
-\omega_{20} R_2&=&\frac{2\omega_1\omega_2}{g\omega\omega_0}
(\epsilon_1 A_1+
\epsilon_2 A_2)\;,
\end{eqnarray}
to which the solution is
\begin{eqnarray}
R_1&=&\frac{\omega_1-\omega_{10}}{\omega_{10}+\omega_1}\;,\nonumber\\
R_2&=&0\;,\nonumber\\
A_1&=&\frac{2\epsilon_2\omega_{10}}{(\epsilon_2-\epsilon_1)
(\omega_{10}+\omega_1)}\;,\nonumber\\
A_2&=&\frac{-2\epsilon_1\omega_{10}}{(\epsilon_2-\epsilon_1)
(\omega_{10}+\omega_1)}\;.
\end{eqnarray}

A similar matching at $z=l$ gives, for the next step in the
procedure, the result below.
As in the linear case some phase factors, and also
some dissipation factors close to one, have been omitted.
\begin{eqnarray}
T_1&=&\frac{4\omega_1\omega_{10}}{(\omega_1+\omega_{10})^2}    \;
(1-i\epsilon_1le^{-i\xi}{\rm sinc}\xi)
\;,\nonumber\\
T_2&=&\frac{2gl\omega_0\omega_{10}(\omega_0-\omega)}
{(\omega_1+\omega_{10})(\omega_2+\omega_{20})}\;
{\rm sinc}\xi
\;,\nonumber\\
A_3&=&\frac{2(\omega_1-\omega_{10})\omega_{10}}{(\omega_1+\omega_{10})^2}\;
(1-i\epsilon_1le^{-i\xi}{\rm sinc}\xi)
\;,\nonumber\\
A_4&=&\frac{gl\omega_0\omega_{10}(\omega_2-\omega_{20})(\omega_0-\omega)}
{\omega_2(\omega_1+\omega_{10})(\omega_2+\omega_{20})}\;
{\rm sinc}\xi
\label{pdcout}\;,
\end{eqnarray}
where we have made use of eq.(\ref{epsprod}), and have put
\begin{equation}
\xi=\frac{(\epsilon_1-\epsilon_2)l}{2}\quad{\rm and}\quad{\rm sinc}\xi=
\frac{\sin\xi}{\xi}\;.
\end{equation}
In an obvious extension of the notation we
used for the linear case, the Poynting vectors
associated with these amplitudes
may be written as
\begin{eqnarray}
\omega_{10}T_1T_1^*&=&\omega_{10}t_{10}^2(1+\gamma)\;,\nonumber\\
\omega_{20}T_2T_2^*&=&\omega_{10}t_{10}t_{20}\gamma
(\omega_0/\omega-1)\;,\nonumber\\
\omega_1A_3A_3^*&=&\omega_{10}t_{10}r_{10}(1+\gamma)\;,\nonumber\\
\omega_2A_4A_4^*&=&\omega_{10}t_{10}r_{20}\gamma
(\omega_0/\omega-1)\;,
\end{eqnarray}
where the new feature, associated with the
down-conversion process, is the coefficient
\begin{equation}
\gamma=\frac{g^2l^2\omega_0^2\omega(\omega_0-\omega)}{
4\omega_1\omega_2}\,{\rm sinc}^2\xi\;.
\end{equation}

As in the linear case, the overall reflection and
transmission coefficients are obtained by simply
adding the intensities associated with each set of
internal reflections. Recalling that, because
the $k_3$ and $k_4$ waves are nondegenerate,
the  factor $\gamma$ is zero for
right-to-left waves, this gives us, to first
order in $\gamma$,
\begin{eqnarray}
r_1&=&r_{10}+r_{10}t_{10}^2(1+\gamma)+r_{10}^3t_{10}^2(1+2\gamma)+\ldots
\nonumber\\ \nonumber\\&=&
\frac{2r_{10}}{1+r_{10}}+\frac{\gamma r_{10}}{(1+r_{10})^2}\;,\\
\nonumber\\
t_1&=&\frac{1-r_{10}}{1+r_{10}}+\frac{\gamma}{(1+r_{10})^2}\;,\\
\nonumber\\
r_2&=&(\omega_0/\omega-1)(
t_{10}\gamma r_{20}t_{20}+t_{10}\gamma r_{20}^3t_{20}+
t_{10}r_{10}^2\gamma r_{20}t_{20}+\ldots)\label{r2}\nonumber\\
\nonumber\\
&=&\frac{(\omega_0-\omega)\gamma r_{20}}{\omega(1+r_{10})(1+r_{20})}\;,\\
\nonumber\\
t_2&=&
\frac{(\omega_0-\omega)\gamma}{\omega(1+r_{10})(1+r_{20})}\;,\label{t2}
\end{eqnarray}
where all of these coefficients have been defined,
as in the linear case, by ratios between the
$z$-components of the Poynting vectors.
We deduce that
\begin{equation}
t_1+r_1-1=\frac{\omega}{\omega_0-\omega}(t_2+r_2)=
\frac{\gamma}{1+r_{10}}\;.\label{photon}
\end{equation}

In the scalar version of
stochastic electrodynamics\cite{mex} the modes of the zeropoint field all have  the
same amplitude, namely $\sqrt{\hbar/2L^3}$; this differs
from the Maxwell version by a factor of $\sqrt\omega$,
the difference being accounted for by the two expressions
for the Poynting vector. Hence the number of
``photons'' in a given mode, including the undetected
half photon of the zeropoint, is proportional to
the modulus-square of that mode's amplitude. So
the above calculation gives, for the number
of idler photons emerging
from unit area of the face $z=l$, and arising
from an incident
wave of zeropoint amplitude and frequency $\omega$,
\begin{equation}
n_i(\omega)=(t_1+r_1-1)/2\;,
\end{equation}
where we have subtracted the zeropoint intensity
in accordance with the theory described in Ref.\cite{pdc4},
while the number of signal photons
is
\begin{equation}
n_s(\omega_0-\omega)=(t_2+r_2)\omega_{10}/(2\omega_{20})\;.
\end{equation}
Now we must refer back to Fig.3. It will be
observed that the total number of ``photons" in
the $\omega$-channel is obtained by adding the
idler photons from one input to the signal photons
from the conjugate input, that is, using eq.(\ref{photon}),
\begin{equation}
n_i(\omega)+n_s(\omega)=\frac{\gamma}{2}\left(\frac{1}{1+r_{10}}+
\frac{1}{1+r_{20}}\,\frac{\cos[\theta(\omega_0-\omega)]}
{\cos[\theta(\omega)]}\right)\;.\label{pdcdet}
\end{equation}
The corresponding output in the other channel is
\begin{equation}
n_i(\omega_0-\omega)+n_s(\omega_0-\omega)=
\frac{\gamma}{2}\left(\frac{1}{1+r_{20}}+
\frac{1}{1+r_{10}}\,\frac{\cos[\theta(\omega)]}
{\cos[\theta(\omega_0-\omega)]}\right)\;.
\end{equation}
Hence the ratio of the photon fluxes is
\begin{equation}
\frac{n_i(\omega)+n_s(\omega)}{n_i(\omega_0-\omega)+n_s(\omega_0-\omega)}=
\frac{\cos[\theta(\omega_0-\omega)]}
{\cos[\theta(\omega)]}\;.
\end{equation}

So we conclude that {\it the photon rate in
a given channel is inversely proportional to
the cosine of the rainbow angle}\footnote
{Actually there are two rainbows --- a forward one
with intensities $(t_1,t_2)$ and a backward one
with $(r_1,r_2)$. This simple relation is between
the {\it sums} of the intensities in the two
rainbows. It becomes rather more complicated
if we confine attention to just the forward
rainbow, but, since that contains about 96
percent of the total intensity, this simple
relation is still almost exact.
}.
In the standard nonlocal theory associated with
Fig.1, by contrast, the above ratio is one. Indeed, it is an
essential part of the energy-conservation
argument that PDC ``photons" must be created in
pairs. In the local, consistently field-theoretic
approach we are advocating here, energy is still
conserved, but the units for energy transactions
are no longer photons. Indeed, although the result
we just obtained was stated in terms of photon
fluxes, these are really just Poynting vectors
with an appropriate zeropoint subtraction.

There seems little chance of finding out
directly which of these theories is correct;
the difference between the two ratios is small,
since the rainbow angles are typically around
10 degrees, and it is not possible to
measure at all accurately the efficiency
of light detectors as a function of
frequency. It is true that some of the
experiments we have analysed, using the
standard theory, in Refs.\cite{pdc1,pdc2,pdc3,pdc4},
have slightly different results in the present
theory, for example the fringe visibility in
the experiment of Zou, Wang and Mandel\cite{zwm}.
Some details will be published shortly, but we
can say that an experimental discrimination will
be very difficult.
\section{Parametric up conversion from the vacuum}
\label{puc}
There is, however, at least one prediction
of the new theory which differs dramatically
from the standard theory. An incident wave
of frequency $\omega$, as well as being
down converted, by the pump, to give
a PDC signal of frequency $\omega_0-\omega$,
may also be {\it up converted to give a
PUC signal} of frequency $\omega_0+\omega$.
We depict this phenomenon, which is well
known\cite{saleh} in classical nonlinear optics,
in Fig.4. 
Note that the angle of incidence, $\theta_u(\omega)$,
at which PUC occurs is quite different from the PDC
angle, which in Fig.2 was denoted simply $\theta(\omega)$,
but which we should now call $\theta_d(\omega)$.

Now, following the same argument which led us from
Fig.2 to Fig.3, we predict the phenomenon of
PUC from the Vacuum, which we depict in Fig.5.

Let us calculate the intensity of this PUC rainbow. There
is an important difference from the PDC situation, arising
from a different relation between the frequency eigenvalues
inside the crystal. Eqn.(\ref{epsprod}) must be replaced by
\begin{equation}
\epsilon_1\epsilon_2=-\frac{g^2\omega(\omega_0+\omega)\omega_0^2}
{4\omega_1\omega_2}\;.
\end{equation}
As a consequence we find that, in contrast with PDC,
the intensity of the PUC idler is less than that of the input,
and eq.(\ref{photon}) must be replaced by
\begin{equation}
1-t_1-r_1=\frac{\omega}{\omega_0+\omega}(t_2+r_2)=
\frac{\gamma}{1+r_{10}}\;.
\end{equation}
It now follows that the ``photon" fluxes in the
two outgoing channels are
\begin{eqnarray}
n_i(\omega)+n_s(\omega)&=&\frac{\gamma}{2}\left(
\frac{1}{1+r_{20}}\,\frac{\cos[\theta(\omega_0+\omega)]}
{\cos[\theta(\omega)]}-\frac{1}{1+r_{10}}\right)\;,\nonumber\\
\label{pucdet}\\
n_i(\omega_0+\omega)+n_s(\omega_0+\omega)&=&
\frac{\gamma}{2}\left(
\frac{1}{1+r_{10}}\,\frac{\cos[\theta(\omega)]}
{\cos[\theta(\omega_0+\omega)]}-\frac{1}{1+r_{20}}\right)\;.\nonumber\\
\end{eqnarray}

The cosines occurring here all differ from one by a
few percent. The reflection coefficients differ from zero
by a few percent, and from each other by a few tenths
of a percent. Since (see Fig.5) the cosine of $\theta(\omega_0+\omega)$
is greater than the cosine of $\theta(\omega)$, it follows that
the photon flux in the $\omega$-channel is positive, while
that in the $(\omega_0+\omega)$-channel is negative,
which means simply that the overall intensity there
is below the zeropoint and that no detection events
will occur. {\it We expect to see detection events
only in the $\omega$-channel.} A comparison between
eqs.(\ref{pucdet}) and (\ref{pdcdet}) shows that,
because of the sign difference, and the closeness
of the cosines to one and of the reflection coefficients
to zero,
the intensity of the PUC rainbow is expected to
be only a few percent of the PDC rainbow. This
may explain why nobody has yet reported seeing it.

We may calculate quite easily the approximate
relative positions of the PDC and PUC rainbows.
In the PDC case the rainbow angle is determined by eq.(\ref{rainbowd}).
For simplicity we shall consider the case
$\omega=\omega_0/2$. Then, defining
\begin{equation}
q_d=\sin^2[\theta_d(\omega_0/2)]\quad,\quad\mu_1=\mu(\omega_0/2)
\quad,\quad\mu_2=\mu(\omega_0)\;,
\end{equation}
eq.(\ref{rainbowd}) tells us that
\begin{equation}
q_d=\mu_1^2-\mu_2^2\;.
\end{equation}
Now, similarly, let us define
\begin{equation}
q_u=\sin^2[\theta_u(\omega_0/2)]\quad,\quad\mu_3=\mu(3\omega_0/2)\;.
\end{equation}
The equation defining the PUC rainbow angle at this
frequency is
\begin{equation}
-\sqrt{\mu_1^2-q_u}+\sqrt{9\mu_3^2-q_u}=2\mu_2\;,
\end{equation}
which gives
\begin{equation}
q_u=\frac{1}{16\mu_2^2}\;
[36\mu_1^2\mu_3^2-(9\mu_3^2-4\mu_2^2+\mu_1^2)^2]\;.\label{qu}
\end{equation}
We have just established that $\mu_1^2=\mu_2^2+q_d$,
so let us make a linear approximation
to $\mu_3$, namely
\begin{equation}
\mu_3^2=\mu_2^2-q_d\;.
\end{equation}
Substituting in eq.(\ref{qu}) this gives us that
\begin{equation}
q_u=6q_d-25q_d^2/4\mu_2^2\;.
\end{equation}
So, if the degenerate PDC mode is at 10 degrees,
then the PUC rainbow has its $\omega_0/2$ mode
at about 25 degrees, which means that the two
rainbows are well separated.

A detailed calculation of the intensity, using
the above approximation, gives that the PUC
intensity at this frequency is 3.3 percent of
the PDC intensity. Naturally this should be
taken as an order-of-magnitude prediction only,
since the whole calculation was based on a
scalar version of the Maxwell equations.

\noindent
{\bf Acknowledgement}

\noindent
I have had a lot of help with the ideas behind this article,
and also in developing the argument, from Emilio Santos.

{\bf Figure captions}
\begin{enumerate}
\item{
PDC --- photon-theoretic version. A laser photon
down converts into a conjugate pair of PDC photons
with conservation of energy.}
\item
{Classical PDC. When a wave of
frequency $\omega$ is incident, at a certain angle
$\theta(\omega)$, on a nonlinear crystal
pumped at frequency $\omega_0$, a signal
of frequency $\omega_0-\omega$ is emitted
in a certain conjugate direction. The modified
input wave is called the idler.}
\item
{PDC from the vacuum --- field-theoretic version. Both of
the outgoing signals are above zeropoint intensity, and
hence give photomultiplier counts.}
\item
{PUC. In contrast with PDC
the output signal has its
transverse component in the same direction as that of
the idler.}
\item
{PUC from the vacuum.  Only one
of the outgoing signals is above the zeropoint intensity. The
other one, depicted by an interrupted line, is below
zeropoint intensity.}
\end{enumerate}
\begin{figure}
\unitlength=0.8mm
\linethickness{0.4pt}
\begin{picture}(159.33,44.67)
\put(70.00,44.67){\line(0,-1){40.00}}
\put(90.00,4.67){\line(0,1){40.00}}
\thicklines
\put(0.00,24.67){\line(1,0){70.00}}
\thinlines
\put(90.00,24.67){\line(5,1){69.33}}
\put(159.33,10.67){\line(-5,1){69.33}}
\put(80.00,24.67){\makebox(0,0)[cc]{{NLC}}}
\put(80.00,31.67){\makebox(0,0)[cc]{{}}}
\put(80.00,17.67){\makebox(0,0)[cc]{{}}}
\put(12.67,24.67){\vector(1,0){7.67}}
\put(119.33,18.67){\vector(4,-1){1.00}}
\put(119.67,30.67){\vector(4,1){1.00}}
\end{picture}
\caption{}
\end{figure}
\begin{figure}
\unitlength=0.8mm
\linethickness{0.4pt}
\begin{picture}(159.33,44.67)
\put(70.00,44.67){\line(0,-1){40.00}}
\put(90.00,4.67){\line(0,1){40.00}}
\thicklines
\put(0.00,24.67){\line(1,0){70.00}}
\thinlines
\put(70.00,24.67){\line(-5,1){70.00}}
\put(90.00,24.67){\line(5,1){69.33}}
\put(159.33,10.67){\line(-5,1){69.33}}
\put(31.33,19.67){\makebox(0,0)[cc]{{\footnotesize laser}}}
\put(31.33,38.67){\makebox(0,0)[cc]{{\footnotesize input$(\omega)$}}}
\put(122.33,38.34){\makebox(0,0)[cc]{{\footnotesize signal$(\omega_0-\omega)$}}}
\put(122.00,10.00){\makebox(0,0)[cc]{{\footnotesize idler$(\omega)$}}}
\put(80.00,24.67){\makebox(0,0)[cc]{NLC}}
\put(80.00,31.67){\makebox(0,0)[cc]{{}}}
\put(80.00,17.67){\makebox(0,0)[cc]{{}}}
\put(12.67,24.67){\vector(1,0){7.67}}
\put(14.67,35.67){\vector(4,-1){1.00}}
\put(119.33,18.67){\vector(4,-1){1.00}}
\put(119.67,30.34){\vector(4,1){1.00}}
\put(32.67,28.00){\makebox(0,0)[cc]{{\footnotesize $\theta(\omega)$}}}
\end{picture}
\caption{}
\end{figure}
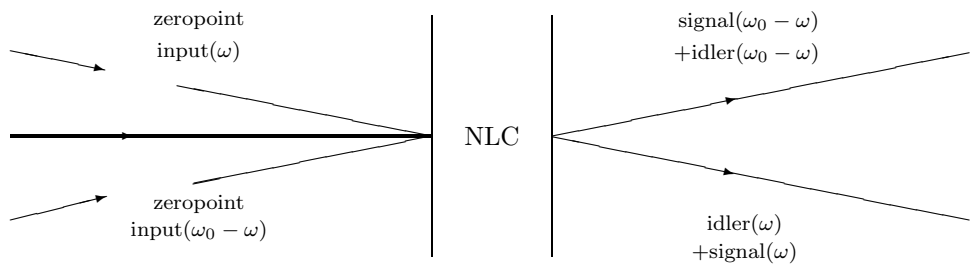
\begin{figure}
\unitlength=0.8mm
\linethickness{0.4pt}
\begin{picture}(159.33,44.67)
\put(70.00,44.67){\line(0,-1){40.00}}
\put(90.00,4.67){\line(0,1){40.00}}
\thicklines
\put(0.00,24.67){\line(1,0){70.00}}
\thinlines
\put(31.67,28.34){\makebox(0,0)[cc]{{}}}
\put(90.00,24.67){\line(5,1){69.33}}
\put(159.33,10.67){\line(-5,1){69.33}}
\put(11.33,19.67){\makebox(0,0)[cc]{}}
\put(31.33,38.67){\makebox(0,0)[cc]{{\footnotesize input($\omega$)}}}
\put(122.33,38.34){\makebox(0,0)[cc]{{\footnotesize +idler($\omega_0-\omega$)}}}
\put(122.00,10.00){\makebox(0,0)[cc]{{\footnotesize idler($\omega$)}}}
\put(80.00,24.67){\makebox(0,0)[cc]{NLC}}
\put(80.00,31.67){\makebox(0,0)[cc]{{}}}
\put(80.00,17.67){\makebox(0,0)[cc]{{}}}
\put(12.67,24.67){\vector(1,0){7.67}}
\put(14.67,36.00){\vector(4,-1){1.00}}
\put(119.33,18.67){\vector(4,-1){1.00}}
\put(119.67,30.67){\vector(4,1){1.00}}
\put(70.00,24.67){\line(-5,1){42.33}}
\put(15.67,35.67){\line(-5,1){15.67}}
\put(0.00,38.13){\line(0,0){0.00}}
\put(31.33,44.00){\makebox(0,0)[cc]{{\footnotesize zeropoint}}}
\put(122.67,43.67){\makebox(0,0)[cc]{{\footnotesize signal($\omega_0-\omega$)}}}
\put(122.00,5.33){\makebox(0,0)[cc]{{\footnotesize +signal($\omega$)}}}
\put(0.00,10.67){\vector(4,1){16.00}}
\put(70.00,24.67){\line(-5,-1){39.67}}
\put(31.33,13.33){\makebox(0,0)[cc]{{\footnotesize zeropoint}}}
\put(31.33,9.00){\makebox(0,0)[cc]{{\footnotesize
input($\omega_0-\omega$)}}}
\put(33.33,21.67){\makebox(0,0)[cc]{{}}}
\end{picture}
\caption{}
\end{figure}
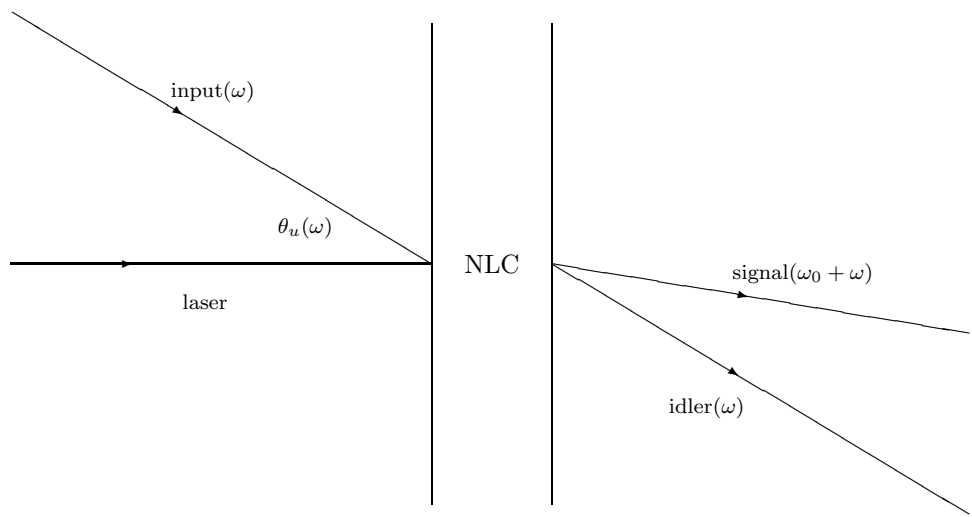
\begin{figure}
\unitlength=0.8mm
\linethickness{0.4pt}
\begin{picture}(159.33,86.80)
\put(70.00,85.00){\line(0,-1){80.00}}
\put(90.00,5.00){\line(0,1){80.00}}
\thicklines
\put(0.00,45.00){\line(1,0){70.00}}
\thinlines
\put(80.00,45.00){\makebox(0,0)[cc]{{NLC}}}
\put(80.00,52.00){\makebox(0,0)[cc]{{}}}
\put(80.00,38.00){\makebox(0,0)[cc]{{}}}
\put(12.67,45.00){\vector(1,0){7.67}}
\put(70.00,45.00){\line(-5,3){69.67}}
\put(90.00,45.00){\line(5,-3){69.33}}
\put(159.33,3.40){\line(0,0){0.00}}
\put(90.00,45.00){\line(6,-1){69.33}}
\put(49.00,51.00){\makebox(0,0)[cc]{{\footnotesize $\theta_u(\omega)$}}}
\put(32.33,38.67){\makebox(0,0)[cc]{{\footnotesize laser}}}
\put(33.67,73.67){\makebox(0,0)[cc]{{\footnotesize input($\omega$)}}}
\put(131.67,43.33){\makebox(0,0)[cc]{{\footnotesize signal($\omega_0+\omega$)}}}
\put(115.67,21.00){\makebox(0,0)[cc]{{\footnotesize idler($\omega$)}}}
\put(120.00,27.00){\vector(3,-2){1.00}}
\put(121.67,39.67){\vector(4,-1){1.00}}
\put(27.67,70.33){\vector(3,-2){1.00}}
\end{picture}
\caption{}
\end{figure}
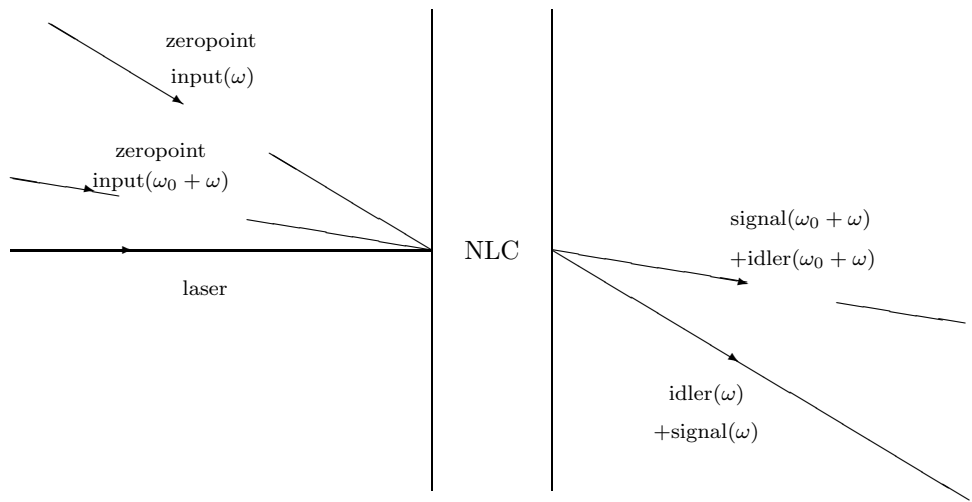
\begin{figure}
\unitlength=0.8mm
\linethickness{0.4pt}
\begin{picture}(159.33,85.00)
\put(70.00,85.00){\line(0,-1){80.00}}
\put(90.00,5.00){\line(0,1){80.00}}
\thicklines
\put(0.00,45.00){\line(1,0){70.00}}
\thinlines
\put(80.00,45.00){\makebox(0,0)[cc]{{NLC}}}
\put(80.00,52.00){\makebox(0,0)[cc]{{}}}
\put(80.00,38.00){\makebox(0,0)[cc]{{}}}
\put(12.67,45.00){\vector(1,0){7.67}}
\put(90.00,45.00){\line(5,-3){69.33}}
\put(159.33,3.40){\line(0,0){0.00}}
\put(32.33,38.67){\makebox(0,0)[cc]{{\footnotesize laser}}}
\put(33.67,73.67){\makebox(0,0)[cc]{{\footnotesize input($\omega$)}}}
\put(131.67,43.33){\makebox(0,0)[cc]{{\footnotesize +idler($\omega_0+\omega$)}}}
\put(115.67,21.00){\makebox(0,0)[cc]{{\footnotesize idler($\omega$)}}}
\put(120.00,27.00){\vector(3,-2){1.00}}
\put(121.67,39.67){\vector(4,-1){1.00}}
\put(27.67,70.00){\vector(3,-2){1.00}}
\put(33.33,79.67){\makebox(0,0)[cc]{{\footnotesize zeropoint}}}
\put(70.00,45.00){\line(-6,1){30.67}}
\put(18.00,54.00){\line(-6,1){18.00}}
\put(13.00,55.00){\vector(4,-1){1.00}}
\put(25.00,56.33){\makebox(0,0)[cc]{{\footnotesize input($\omega_0+\omega$)}}}
\put(25.00,61.33){\makebox(0,0)[cc]{{\footnotesize zeropoint}}}
\put(131.33,50.00){\makebox(0,0)[cc]{{\footnotesize signal($\omega_0+\omega$)}}}
\put(115.67,14.67){\makebox(0,0)[cc]{{\footnotesize +signal($\omega$)}}}
\put(70.00,45.00){\line(-5,3){27.00}}
\put(28.67,69.33){\line(-5,3){22.33}}
\put(90.00,45.00){\line(6,-1){32.33}}
\put(137.33,36.33){\line(6,-1){21.33}}
\end{picture}
\caption{}
\end{figure}
\end{document}